\documentclass{ifacconf}

\usepackage{graphicx}      
\usepackage{natbib}        
\usepackage{amsmath,amsfonts}
\usepackage{xcolor}

\newcommand{\R}{\mathbb{R}}
\newcommand{\Pc}{\mathcal{P}}
\DeclareMathOperator{\He}{He}

\begin{document}
\begin{frontmatter}

\title{A Comparison of LPV Gain Scheduling and Control Contraction Metrics for Nonlinear Control \thanksref{footnoteinfo}} 

\thanks[footnoteinfo]{This work was supported by the Australian Research Council. This work has received funding from the European Research Council (ERC) under the European Union’s Horizon 2020 research and innovation programme (grant agreement nr. 714663).}

\author[ACFR]{Ruigang Wang} 
\author[TUE]{Roland T\'{o}th} 
\author[ACFR]{Ian R. Manchester} 

\address[ACFR]{Australian Centre for Field Robotics, The University of Sydney, NSW 2006, Australia (e-mail: \{ruigang.wang, ian.manchester\}@sydney.edu.au).}
\address[TUE]{Department of Electrical Engineering, Eindhoven University of Technology, P.O. Box 513, 5600 MB Eindhoven, The Netherlands (e-mail: r.toth@tue.nl).}

\begin{abstract}                
Gain-scheduled control based on linear parameter-varying (LPV) models derived from local linearizations is a widespread nonlinear technique for tracking time-varying setpoints. Recently, a nonlinear control scheme based on Control Contraction Metrics (CCMs) has been developed to track arbitrary admissible trajectories. This paper presents a comparison study of these two approaches. We show that the CCM based approach is an extended gain-scheduled control scheme which achieves global reference-independent stability and performance through an exact control realization which integrates a series of local LPV controllers on a particular path between the current and reference states.
\end{abstract}


\end{frontmatter}

\section{Introduction}

In many industrial applications, systems with nonlinear dynamical behavior are required to be operated in a wide range of operating conditions. A widespread approach for this situation is \emph{gain-scheduled control} using \emph{linear parameter-varying} (LPV) system representations \citep{Papageorgiou:2000, Rugh:2000, Klatt:1998}. The underlying idea is to introduce a so-called scheduling variable $\sigma$ that indicates the current operating point of the system and construct a \emph{linear} model that describes the local, linearized dynamics of the plant around each point. The parameters of the resulting model varies with $\sigma$. Next, assuming that $\sigma$ is an external variable (independent from the inputs) an LPV controller dependent on $\sigma$ is designed that, by using linear system theory \citep{Becker:1994}, ensures stability and performance specifications for the LPV model under possible variations of $\sigma$ in a user specified region of operating conditions $\mathcal{P}$.  Finally, a nonlinear control law is obtained by substituting $\sigma$ with measured information of the operating point of the system. 

There are many approaches available to construct an LPV model of the plant based on this methodology, see \cite{Toth:2013} for an overview. Typically, the plant is linearized around a given set of equilibrium points (griding of $\mathcal{P}$) and the resulting set of LTI models are interpolated over $\mathcal{P}$ or linearization is accomplished over input and state trajectories. Similarly, the LPV controller can be obtained by designing LTI controllers separately for finite set of values of $\sigma$ and then interpolating these LTI controllers on $\mathcal{P}$ or parametrizing an LPV controller and solving the stabilization and performance problem jointly over $\mathcal{P}$. Typically, local equilibrium-independent stability and performance can be achieved via these methods, requiring $\sigma$ to be ``sufficiently slow-varying" \citep{Rugh:2000}. 

It is important to highlight that next to gain-scheduling based modeling and control, which is often called a local LPV approach, modern LPV control methods are based on directly transforming the nonlinear system model via the so-called \emph{global embedding principle}, see \cite{Toth:2010,Hoffmann:2015} for an overview, and then synthesizing an LPV control law that gives stability and performance guarantees over all possible variations of $\sigma$. Such methods had been thought superior over gain-scheduling techniques as they provided direct stability guarantees for the embedded nonlinear system following a differential inclusion concept. However, recent studies \cite{Scorletti:2015} indicate that performance issues for reference tracking objectives might still be present. To address this problem, a strong notion - incremental stability was considered, and the corresponding LPV modeling has a connection to the local linearization of the plant.

Contraction theory which builds on global stability results from local analysis has gained much attention for nonlinear systems \citep{Lohmiller:1998, Forni:2014}. Related works include velocity linearization \citep{Leith:2000} and  G\^{a}teaux derivative \citep{Fromion:2001,Fromion:2003}. Recently, contraction analysis was extended to constructive nonlinear control design by using a differential version of control Lyapunov function - Control Contraction Metric (CCM) \citep{Manchester:2017,Manchester:2018}. Further extensions of the CCM based approach include distributed control \citep{Shiromoto:2018} and distributed economic model predictive control (MPC) \citep{Wang:2017}.

The main contribution of this paper is a comparison study between the CCM based nonlinear control approach and the LPV gain scheduling technique using local linearization. For simplicity, only state-feedback control design is considered. We show that CCM based control is an extended LPV gain scheduling approach. First, the so-called \emph{differential dynamics} in contraction theory can be seen as a local LPV model which takes linearization along any admissible solution rather than an equilibrium family in conventional gain-scheduling. Second, similar parameter-dependent linear matrix inequality (LMI) conditions are derived as in local LPV synthesis. One difference is that the CCM based approach explicitly takes the original nonlinear plant into account leading to less conservative results. Furthermore, the control realization integrates a series of local controllers on a particular path joining the current and reference state trajectory, which leads to an \emph{exact} realization without any hidden coupling term. Based on this, local stability and performance design can be carried onto the entire state space as the length of the path shrinks exponentially. 

{\it Paper outline.} Section~\ref{sec:preliminaries} presents formulations of different stability and performance. Section~\ref{sec:gain-scheduling} gives a brief review of the LPV gain scheduling approach using local linearization, which is mainly adopted from \cite{Rugh:2000}. Section~\ref{sec:contraction} discusses the various connections and extensions between CCM and LPV based approaches. An illustrative example is presented in Section~\ref{sec:case-study}.

\section{Preliminaries} \label{sec:preliminaries}

Let $ |x| $ be the Euclidean norm of a vector $ x $. For any matrix $ A\in \R^{n\times n} $, we use the notation $ \He\{A\}:=A+A^\top $. Positive (negative) definiteness of a Hermitian matrix $X$ is denoted as $X\succ 0$ ($X\prec 0$). $ \mathcal{C}^k $ denotes the set of vector signals on $ \R $ which are $ k^\mathrm{th}$ times differentiable. $ \mathcal{L}_2 $ is the space of square-integrable vector signals on $ \R_{\geq 0} $, i.e., $ \|f\|:=\sqrt{\int_{0}^{\infty}|f(t)|^2dt}<\infty $. The causal truncation $ (\cdot)_T $ is defined by $ (f)_T(t):=f(t) $ for $ t\leq T $ and 0 otherwise. $ \mathcal{L}_2^\mathrm{e} $ is the space of vector signals on $ \R_{\geq 0} $ whose causal truncation belongs to $ \mathcal{L}_2 $.

In this paper, we consider a nonlinear system
\begin{equation}\label{eq:system}
\dot{x}=f(x,u,w),\; z=h(x,u,w)
\end{equation}
where $ x(t)\in\R^{n_x},\, u(t)\in\R^{n_u},\,w(t)\in\R^{n_w},\, z(t)\in\R^{n_z} $ are state, control, external input and performance output signals at time $ t\in\R_{\geq 0} $, respectively. The functions $ f$ and $h$ are assumed to be smooth and time-invariant. We define a target trajectory to be a forward-complete solution of \eqref{eq:system}, i.e., a pair $ (x^*,u^*,w^*,z^*)(\cdot) $ with $ x^*(\cdot) $ piecewise differentiable and $ (u^*,w^*,z^*)(\cdot) $ piecewise continuous satisfying \eqref{eq:system} for all $ t\in\R_{\geq 0} $. The target trajectory is said to be an equilibrium if $ (x^*,u^*,w^*,z^*)(\cdot)=(x_\mathrm{e},u_\mathrm{e},w_\mathrm{e},z_\mathrm{e}) $. For simplicity, we assume that the nominal external input is $ w^*(\cdot)=0 $. We will consider state-feedback controllers of the form
\begin{equation}\label{eq:control}
u(t)=\kappa(x(t),x^*(t),u^*(t)).
\end{equation}

To define a nonlinear control problem precisely, we must be specific about the notion of stability and performance. The closed-loop system of \eqref{eq:system} and \eqref{eq:control} is said to be globally asymptotically stable with respect to the target trajectory $ (x^*,u^*)(\cdot)$ if the closed-loop solution $ x(t) $ of the nominal system (i.e., $ w=w^* $) exists and satisfies
\begin{itemize}
	\item[1)] For any $ \epsilon $ there exists an $ \rho $ such that $ |x(0)-x^*(0)|<\rho $ implies $ |x(t)-x^*(t)|<\epsilon $,
	\item[2)] For any initial condition $ x(0)\in\R^n $, the closed-loop solution satisfies $ |x(t)-x^*(t)|\rightarrow 0 $.
\end{itemize}  
Global exponential stability is a stronger notion and requires that there exists a $ R $ and $ \lambda $ such that
\begin{equation}\label{eq:expo-stability}
	|x(t)-x^*(t)|\leq Re^{-\lambda t}|x(0)-x^*(0)|,\quad \forall x(0)\in\R^n.
\end{equation}
The closed-loop system is said to achieve $ \mathcal{L}_2 $-gain performance of $ \alpha $ for the target trajectory $ (x^*,u^*,w^*,z^*)(\cdot) $, if for any initial condition $ x(0) $ and input $ w $ such that $ w-w^*\in\mathcal{L}_2^e $, and for all $ T>0 $ solutions exist and satisfy
\begin{equation}\label{eq:L2-gain}
	\|(z-z^*)_T\|^2\leq \alpha^2\|(w-w^*)_T\|+\beta(x(0),x^*(0))
\end{equation}
where $ \beta(x_1,x_2)\geq 0 $ with $ \beta(x,x)=0 $.

The above reference-related stability and performance notions allow us to investigate different formulations of control objectives in a unified way. Let $ \mathfrak{B}^* $ be the set of target trajectories which are used for the definitions of asymptotic (exponential) stability and $ \mathcal{L}_2 $ performance. In a standard formulation, $ \mathfrak{B}^* $ only contains one ``preferred'' trajectory -- the zero solution (i.e., $ f(0,0,0)=0,\,h(0,0,0)=0 $). A stronger formulation,  called \emph{equilibrium-independent asymptotic (exponential) stability and $ \mathcal{L}_2 $ gain}, is referred to the case where $ \mathfrak{B}^* $ is chosen to be the set of all possible equilibrium points \citep{Simpson-Porco:2019}. The so-called \emph{universal exponential stability and $ \mathcal{L}_2 $ gain} is an even stronger formulation which requires $ \mathfrak{B}^* $ to include all admissible trajectories of the nominal system (\cite{Manchester:2017,Manchester:2018}). The \emph{incremental} formulation is referred to the case where asymptotic (exponential) stability and $ \mathcal{L}_2 $ gain are satisfied for any pair of system trajectories. Note that universal exponential stability is equivalent to incremental exponential stability. But universal $ \mathcal{L}_2 $ gain is weaker than the incremental one \citep{Manchester:2018}.

\section{Gain Scheduling approach}\label{sec:gain-scheduling} 

\subsection{System Linearization}

To construct an LPV representation for \eqref{eq:system} using local modelling concept, we assume that the equilibrium points are uniquely characterized by $x_\mathrm{e}$. Hence, to describe the equilibrium points associated local dynamics of the system, we introduce a scheduling variable $ \sigma\in\R^{n_\sigma} $ that depends on the state, i.e., 
\begin{equation}\label{eq:schedule-variable}
\sigma=g(x)
\end{equation}
where $g$ is a smooth vector function. Note that $ \sigma $ can also depend on $ w $ if it is measurable. Here the possible trajectories of the scheduling signal $ \sigma(t) $ are assumed to belong to the set
\begin{equation}\label{eq:parameter-description}
\mathcal{T}=\{\sigma\in\mathcal{C}^1: \sigma(t)\in \Pc,\; \dot{\sigma}(t)\in \dot{\Pc},\; \forall t\geq 0\}
\end{equation}
where $ \Pc=\{\sigma\in\R^{n_\sigma}:|\sigma_i|\leq \overline{\sigma}_i\}$ and $\dot{\Pc}=\{p\in\R^{n_\sigma}:  |p_i|\leq \overline{p}_i \} $ with $ 1\leq i\leq n_\sigma $. Using $ \sigma $, the equilibrium family is characterized as the set $ \{(x_\mathrm{e},u_\mathrm{e},z_\mathrm{e},w_\mathrm{e})(\sigma) \}_{\sigma\in\Pc} $ where $ x_\mathrm{e}(\cdot),u_\mathrm{e}(\cdot),z_\mathrm{e}(\cdot),$ $w_\mathrm{e}(\cdot)$ are smooth functions, and $ (x_\mathrm{e},u_\mathrm{e},z_\mathrm{e},w_\mathrm{e})(\sigma) $ is an equilibrium of \eqref{eq:system} for all $ \sigma\in\Pc $. 

By linearizing \eqref{eq:system} around the equilibrium family, we obtain an LPV model as follows:
\begin{equation}\label{eq:LPV-system}
\begin{bmatrix}
\dot{x}_\delta \\ z_\delta
\end{bmatrix}=
\begin{bmatrix}
A(\sigma) & B_\mathrm{u}(\sigma) & B_\mathrm{w}(\sigma) \\
C(\sigma) & D_\mathrm{u}(\sigma) & D_\mathrm{w}(\sigma)
\end{bmatrix}
\begin{bmatrix}
x_\delta \\ u_\delta \\ w_\delta
\end{bmatrix},\; \sigma\in\Pc
\end{equation}
where $x_\delta=x-x_\mathrm{e}(\sigma)$,  $u_\delta=u-u_\mathrm{e}(\sigma)$,	$w_\delta=w-w_\mathrm{e}(\sigma)$,  $z_\delta=z-z_\mathrm{e}(\sigma)$
are \emph{deviation variables}. The matrices $ A,B_\mathrm{u},B_\mathrm{w},C,D_\mathrm{u},D_\mathrm{w} $ are defined as the evaluations of $ \frac{\partial f}{\partial x},\frac{\partial f}{\partial u},\frac{\partial f}{\partial w},\frac{\partial h}{\partial x},\frac{\partial h}{\partial u},\frac{\partial h}{\partial w} $ at the $\sigma$ defined equilibrium point.    

\subsection{Control Synthesis}

Consider the static LPV controller of the form
\begin{equation}\label{eq:LPV-control}
u_\delta=K(\sigma)x_\delta,
\end{equation}
which yields a closed-loop LPV system 
\begin{equation}\label{eq:LPV-CL}
\begin{bmatrix}
\dot{x}_\delta \\ z_\delta
\end{bmatrix}=
\begin{bmatrix}
\mathcal{A}(\sigma) & \mathcal{B}(\sigma) \\
\mathcal{C}(\sigma) & \mathcal{D}(\sigma)
\end{bmatrix}
\begin{bmatrix}
x_\delta \\ w_\delta
\end{bmatrix}
\end{equation}
with $\mathcal{A}=A+B_\mathrm{u}K$, $\mathcal{B}=B_\mathrm{w}$, $\mathcal{C}=C+D_\mathrm{u}K$, $\mathcal{D}=D_\mathrm{w}$. 
\begin{thm}\label{thm:LPV-control} 
	The unforced closed-loop system
	\begin{equation}\label{eq:LPV-unforced-system}
	\dot{x}_\delta=\mathcal{A}(\sigma)x_\delta,\; \sigma\in\mathcal{T}
	\end{equation}
	is exponentially stable if there exists a $ M(\sigma)\succ 0 $ such that
	\begin{equation}\label{eq:LPV-LMI}
	\He\{M(\sigma)\mathcal{A}(\sigma)\}+2\lambda M(\sigma)+\sum_{i=1}^{n_\sigma}\rho_i\frac{\partial M(\sigma)}{\partial \sigma_i} \prec 0
	\end{equation}
	for all $ \sigma\in\Pc $ and $ \rho\in\dot{\Pc} $.
\end{thm}
The above theorem implies that $ V(x_\delta,\sigma):=x_\delta^\top M(\sigma)x_\delta $ is a parameter-dependent Lyapunov function for system \eqref{eq:LPV-unforced-system}. By applying a congruence transformation
\begin{equation}
	W(\sigma)=M^{-1}(\sigma),\quad L(\sigma)=K(\sigma)W(\sigma),
\end{equation}
we can obtain a convex formulation:
\begin{equation}\label{eq:LPV-synthesis}
\He\{A(\sigma)W(\sigma)+B_\mathrm{u}(\sigma)L(\sigma)\}+2\lambda W(\sigma)-\sum_{i=1}^{n_\sigma}\rho_i\frac{\partial W(\sigma)}{\partial \sigma_i} \prec 0
\end{equation}
for all $ \sigma\in\Pc $ and $ \rho\in \dot{\Pc} $. 

Note that due to linearity of the LPV description \eqref{eq:LPV-CL}, exponential stability and $ \mathcal{L}_2 $ gain bound of \eqref{eq:LPV-CL} w.r.t. the origin is equivalent to the equilibrium-independent asymptotic stability and $ \mathcal{L}_2 $ gain \citep{Rugh:2000,Hoffmann:2015}. 
\begin{thm}\label{thm:LPV-performance}
	A controller \eqref{eq:LPV-control} achieves an $ \mathcal{L}_2 $ performance level of $ \alpha $ for LPV system \eqref{eq:LPV-system} if there exists $ M(\sigma)\succ0 $ such that, for all $ \sigma\in\Pc $ and $ \rho\in\dot{\Pc} $,
	\begin{equation}\label{eq:LPV-performance}
	\begin{bmatrix}
	\mathcal{M}(\sigma,\rho) & M(\sigma)\mathcal{B}(\sigma) & \alpha^{-1} \mathcal{C}^\top (\sigma) \\
	\mathcal{B}^\top (\sigma)M(\sigma) & -I & \alpha^{-1} \mathcal{D}^\top (\sigma) \\
	\alpha^{-1} \mathcal{C}(\sigma) & \alpha^{-1} \mathcal{D}(\sigma) & -I
	\end{bmatrix}\prec 0
	\end{equation} 
	where $ \mathcal{M}(\sigma,\rho)=\He\{M(\sigma)\mathcal{A}(\sigma)\}+\sum_{i=1}^{n_\sigma}\rho_i\frac{\partial M(\sigma)}{\partial \sigma_i} $.
\end{thm} 
Based on the above condition, we can synthesize an LPV controller which achieves the minimal $ \mathcal{L}_2 $-gain bound for the closed-loop LPV system. 

\subsection{Controller Realization}\label{sec:lpv-realization}

The LPV control realization problem is to construct a gain-scheduled law $ u=\kappa(x,\sigma) $ such that
\begin{subequations}\label{eq:GS-cond}
	\begin{align}
	u_\mathrm{e}(\sigma)=\kappa(x_\mathrm{e}(\sigma),\sigma), \label{eq:GS-cond-1}\\
	\frac{\partial \kappa}{\partial x}(x_\mathrm{e}(\sigma),\sigma)=K(\sigma). \label{eq:GS-cond-2}
	\end{align}
\end{subequations}
Condition \eqref{eq:GS-cond-2} implies that linearization of $ u=\kappa(x,\sigma) $ at this equilibrium is the LPV controller \eqref{eq:LPV-control}. An intuitive choice of control realization in the literature is 
\begin{equation} \label{eq:GS-control}
u=u_\mathrm{e}(\sigma)+K(\sigma)[x-x_\mathrm{e}(\sigma)].
\end{equation}
Under the assumption that the equilibrium points of \eqref{eq:system} are uniquely characterized by $x_\mathrm{e}$, $\sigma$ can be expressed in terms of $x$ via \eqref{eq:schedule-variable}. Using this relation, \eqref{eq:GS-control} reads as
\begin{equation}\label{eq:GS-control-implementation}
u=u_\mathrm{e}(g(x))+K(g(x))[x-x_\mathrm{e}(g(x))].
\end{equation}
The main ``trick" behind of this gain-scheduling approach is that $\sigma$ is treated as a parametric/dynamic uncertainty throughout the design process, but during controller realization is substituted by a function of a measured variable characterizing the operating point changes (\cite{Rugh:2000}). Although $ \sigma $ is implicitly involved via equilibrium parameterizations, linearization of \eqref{eq:GS-control-implementation} may not satisfy condition \eqref{eq:GS-cond-2} since 
\begin{equation*}
\begin{split}
u_\delta=K(\sigma)x_\delta 
+\left[\frac{\partial u_\mathrm{e}(\sigma)}{\partial\sigma}-K(\sigma)\frac{\partial x_\mathrm{e}(\sigma)}{\partial \sigma}\right]\frac{\partial g}{\partial x}(x_\mathrm{e}(\sigma))x_\delta
\end{split}
\end{equation*}
contains additional terms, called \emph{hidden coupling terms}, compared with \eqref{eq:LPV-control}. These terms may lead to closed-loop instability regardless the fact that exponential stability is achieved in the control synthesis stage, which is a well-known drawback of the local LPV controller (see Example~8 in \cite{Rugh:2000}). 

\subsection{Stability and Performance Assessment}

\begin{figure}[!bt]
	\centering
	\includegraphics[width=0.45\linewidth]{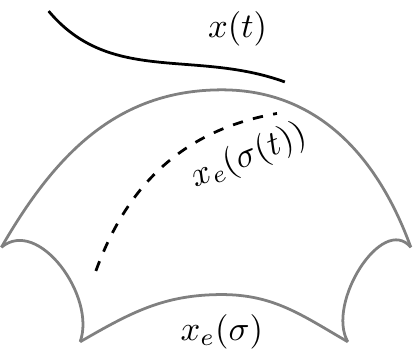}
	\caption{Illustration of gain-scheduling control.}\label{fig:manifold}
\end{figure}

The core idea of gain-scheduled control \eqref{eq:GS-control-implementation} is to track a reference $ x_\mathrm{e}(\sigma(t)) $ lying on the equilibrium manifold, as shown in Fig.~\ref{fig:manifold}. This strategy achieves local equilibrium-independent stability if the schedule signal $ \sigma(t) $ is sufficiently ``slowly varying'' \citep{Rugh:2000}. The main reason is that the scheduled reference trajectory $ (x_\mathrm{e},u_\mathrm{e},w_\mathrm{e},z_\mathrm{e})(\sigma(t)) $ is not admissible to the closed-loop system $ \dot{x}=f(x,\kappa(x,\sigma),w) $ with $\sigma=g(x)$ since simple substitution yields a residual term $ E(\sigma)\dot{\sigma}(t) $ with $ E(\sigma)=\frac{\partial x_\mathrm{e}(\sigma)}{\partial \sigma} $. Therefore, the actual linearization of the closed-loop system with  $ w_\delta(t)=0 $ is
\begin{equation}\label{eq:LPV-actual-linearization}
\dot{x}_\delta=\mathcal{A}(\sigma)x_\delta-E(\sigma)\dot{\sigma}.
\end{equation}
If the rates of parameter variation are not ``sufficiently slow'', the residual terms can drive the state away from the small neighborhood of $ x_\mathrm{e}(\sigma) $, which may violate the local stability design. To ensure global stability and performance, excessive simulations or even experiments are needed.

\section{CCM based Nonlinear Control}\label{sec:contraction}
In this section, we will show the connections and differences between the CCM-based control and local LPV-based gain-scheduled control. Both approaches use very close LPV modeling and control synthesis formulations. The major differences come from the ways to interpret and use the LPV tools. In the CCM-based approach, LPV modeling, design and realization are carried out for the entire state space, which is an extension to the local gain-scheduling approach as it only considers the equilibrium manifold (as shown in Fig.~\ref{fig:manifold}). With this conceptual innovation, the CCM-based approach can provide an exact control realization which does not contain any hidden coupling term and achieve universal  stability and  $ \mathcal{L}_2 $ performance.


Firstly, we recall some basic facts of Riemannian geometry from \cite{Do-Carmo:1992}. A Riemannian metric on $ \R^n $ is a smooth matrix function $ M(x)\succ 0 $ which defines an inner product $ \langle\delta_1,\delta_2 \rangle_x=\delta_1^\top M(x)\delta_2 $ for any two tangent vector $ \delta_1,\delta_2 $. A metric is called \emph{uniformly bounded} if there exist positive constants $ a_2\geq a_1 $ such that $ a_1I\prec M(x)\prec a_2I,\,\forall x\in\R^n $. $ \Gamma(x_0,x_1) $ denotes the set of piecewise smooth paths $ c:[0,1]\rightarrow\R^n $ with $ c(0)=x_0 $ and $ c(1)=x_1 $. The curved length and energy of $ c(\cdot) $ is defined by
\begin{equation*}
\begin{split}
\ell(c)=\int_{0}^{1}\sqrt{\langle c_s,c_s \rangle_{c(s)}}ds\; \text{ and } \;
\varepsilon(c)=\int_{0}^{1}\langle c_s,c_s \rangle_{c(s)}ds
\end{split}
\end{equation*}
where $ c_s=\partial c/\partial s $, respectively. The \emph{geodesic} $ \gamma(\cdot) $ denotes a path with the minimal length, i.e., $ \ell(\gamma)=\inf_{c\in\Gamma(x_0,x_1)}\ell(c) $. The Riemann distance and energy between $ x_0 $ and $ x_1 $ are defined by $ d(x_0,x_1)=\ell(\gamma) $ and $ \varepsilon(x_0,x_1)=\varepsilon(\gamma)=\ell^2(\gamma) $. 

\subsection{System Linearization}

By choosing $ \sigma=(x,u,w) $, we can construct a continuously linearized system (so-called \emph{differential dynamics}) 
\begin{equation}\label{eq:differential-dynamics}
\begin{bmatrix}
\dot{\delta}_x \\ \delta_z
\end{bmatrix}=
\begin{bmatrix}
A(\sigma) & B_u(\sigma) & B_w(\sigma) \\
C(\sigma) & D_u(\sigma) & D_w(\sigma)
\end{bmatrix}
\begin{bmatrix}
\delta_x \\ \delta_u \\ \delta_w
\end{bmatrix}
\end{equation}
where the matrices $ A, B_u, B_w, C, D_u, D_w $ are defined in a similar way as in the local gain-scheduling approach. The variables $ \delta_x,\delta_u,\delta_w,\delta_z $ are the \emph{virtual displacement} between neighboring solutions \citep{Lohmiller:1998} or the \emph{tangent vector} of the solution manifold \citep{Forni:2014}. Other related linearization techniques include velocity linearization \citep{Leith:2000} and G\^{a}teaux derivative \citep{Fromion:2001}.
\begin{rem}
Note that the differential dynamics \eqref{eq:differential-dynamics} can be seen as a local LPV system defined on the entire state space rather than the equilibrium manifold.
\end{rem}
\subsection{Control Synthesis}

The control synthesis searches for a differential controller:
\begin{equation}\label{eq:local-controller}
\delta_u=K(\sigma)\delta_x:=K(x,u)\delta_x
\end{equation}
which stabilize the unforced closed-loop dynamics 
\begin{equation}\label{eq:differential-CL}
\dot{\delta}_x=\mathcal{A}(\sigma)\delta_x:=[A(\sigma)+B_u(\sigma)K(\sigma)]\delta_x.
\end{equation}
It can be achieved by a sufficient condition as follows
\begin{equation}\label{eq:cond-ccm}
	\He\{M(x)\mathcal{A}(\sigma)\}+2\lambda M(x)+\sum_{i=1}^{n_x}f_i(\sigma) \frac{\partial M(x)}{\partial x_i} \prec 0
\end{equation}
where $ f_i $ is the model of the state $ x_i $. The above synthesis formulation is very close to \eqref{eq:LPV-LMI}. Thus, similar convexation technique can be applied here. The main difference between these two formulation is that \eqref{eq:cond-ccm} uses detailed model \eqref{eq:system} to describe the scheduling signal $ \sigma(t) $ while \eqref{eq:LPV-synthesis} uses coarse description \eqref{eq:parameter-description} which only contains the region of the parameter and its variation. This can lead to less conservative results. For instance, a non-uniform metric $ M(x) $ can be found even if both the parameter $ \sigma $ and its variation $ \dot{\sigma} $ are unbounded, e.g., the system $ \dot{x}=-x-x^3 $ admits a non-uniform metric $ M(x)=1+3x^2 $.

Here $ V(x,\delta_x)=\delta_x^\top M(x)\delta_x $ is called control contraction metric (CCM), which is a differential control Lyapunov function validated everywhere in the state space. Since the schedule variable $ \sigma $ in the gain scheduling approach only depends on $ x $, the control Lyapunov matrix $ M(\sigma) $ obtained from \eqref{eq:LPV-LMI} can be seen as a CCM defined on the equilibrium manifold.

For performance analysis, we can obtain a formulation similar to \eqref{eq:LPV-performance}:
\begin{equation}\label{eq:contraction-performance-LMI}
\begin{bmatrix}
\mathcal{M} & M\mathcal{B} & \alpha^{-1} \mathcal{C}^\top \\
\mathcal{B}^\top M & -I & \alpha^{-1} \mathcal{D}^\top \\
\alpha^{-1} \mathcal{C} & \alpha^{-1} \mathcal{D} & -I
\end{bmatrix}\prec 0
\end{equation}  
where $ \mathcal{M}=\He\{M(x)\mathcal{A}(\sigma)\}+\sum_{i=1}^{n_x}f_i(\sigma)\frac{\partial M(x)}{\partial x_i} $.

\subsection{Controller Realization}
As discussed in Section~\ref{sec:lpv-realization}, the differential controller \eqref{eq:local-controller} is generally not completely integrable, i.e., there does not exist a gain-scheduled law $ \kappa(\cdot) $ whose Jacobian is $ K $. Unlike the LPV control, the CCM based approach only considers a much weaker condition - the path-integrability of $ K $.

Let $ \gamma $ be a geodesic (a minimal path) joining $ x^*(t) $ and $ x(t) $, which can be obtained by solving a simple model predictive control problem online \citep{Leung:2017}. The state-feedback law is the integral of the differential control \eqref{eq:local-controller} over the path $ \gamma $:
\begin{equation}\label{eq:contraction-control}
	u(t)=\kappa(x(t),x^*(t)):=\kappa_\gamma(t,1)
\end{equation} 
where $ \kappa_\gamma(t,s) $ is the unique solution of the following integral equation
\begin{equation}\label{eq:path-integral}
\kappa_\gamma(t,s):=u^*(t)+\int_0^sK(\gamma(\mathfrak{s}),\kappa_\gamma(t,\mathfrak{s}))\gamma_s(\mathfrak{s}) d\mathfrak{s}.
\end{equation}
\begin{rem}
	Note that $ \kappa_\gamma(t,s) $ satisfies $ \kappa_\gamma(t,0)=u^* $ and
	\[
	\frac{\partial \kappa_\gamma(t,s)}{\partial s}=K(\gamma(s),\kappa_\gamma(t,s))\gamma_s(s),\;\forall s\in [0,1].
	\] 
	Thus, it does not contain any hidden coupling term and serves as an \emph{exact} realization for the differential controller \eqref{eq:local-controller}. Moreover, it can be also applied to those approaches using incremental analysis \citep{Fromion:2003} where exact realization for general nonlinear systems is an open problem \citep{Scorletti:2015}.  
\end{rem}
 
\begin{figure}[!bt]
	\centering
	\includegraphics[width=0.5\linewidth]{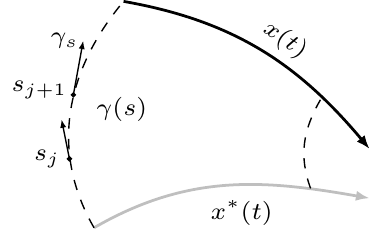}
	\caption{Geometric illustration of the control realization in CCM based approach.} \label{fig:path-integral}
\end{figure}

We give a geometric interpretation about the interconnection between the path of control signal $ \kappa_\gamma(t,s) $ and the local LPV controller \eqref{eq:GS-control}. Let $ 0=s_0<s_1<\cdots<s_N=1 $ with $ s_{j+1}-s_j $ be sufficiently small. For any frozen time $ t $, the integral equation \eqref{eq:path-integral} gives 
\begin{equation*}
\kappa_\gamma(s_{j+1})\approx \kappa_\gamma(s_{j})+K(\gamma(s_j),\kappa_\gamma(s_{j}),w)[\gamma(s_{j+1})-\gamma(s_j)],
\end{equation*}
where the argument $ t $ is omitted for simplicity. Thus, $ \kappa_\gamma(s_{j+1}) $ is an LPV controller \eqref{eq:GS-control} that stabilizes the state $ \gamma(s_{j+1}) $ around $ \gamma(s_j) $, as shown in Fig.~\ref{fig:path-integral}. Based on this observation, $ \kappa_\gamma $ integrates a series of local LPV controllers \eqref{eq:local-controller} along a particular path $ \gamma $ and the CCM based gain scheduling law \eqref{eq:contraction-control} is the corresponding control action to the measured state  $ x(t)=\gamma(t,1) $.
 
\subsection{Stability and Performance Assessment}

Since the control realization \eqref{eq:path-integral} is exact, the differential dynamics of the closed-loop system along the path $ \gamma $ is same as \eqref{eq:differential-CL}, which is exponentially stable:
\begin{equation*}
\begin{split}
\frac{d}{dt}(\delta_x^\top M(x)\delta_x)=\delta_x^\top\dot{M}\delta_x+2\delta_x^\top M\mathcal{A}\delta_x  \leq -2\lambda\delta_x^\top M(x)\delta_x.
\end{split}
\end{equation*}
The global stability follows by integrating the above inequality along a geodesic $ \gamma $: 
\begin{equation}\label{eq:convergence}
\frac{d}{dt}\varepsilon(x^*(t),x(t))\leq -2\lambda\varepsilon(x^*(t),x(t)).
\end{equation}
An explanation using LPV concepts is given as follows. The smooth path $ \gamma $ can be understood as a ``chain'' of many states joining the current state $ x(t) $ to the reference point $ x^*(t) $, and the tangent vector $ \delta_x=\gamma_s $ as a ``link'' whose behavior is described by the closed-loop differential dynamics \eqref{eq:differential-CL}, as shown in Fig.~\ref{fig:path-integral}. The convergence of $ x(t) $ to $ x^*(t) $ can be inferred since each link in the chain gets shorter due to local stability. The following theorems give the global stability and performance results for the CCM based approach.
\begin{thm}[\cite{Manchester:2017}] 
	If there exists a uniformly bounded metric $ a_1I\leq M(x)\leq a_2I $ for which \eqref{eq:cond-ccm} holds for all $ x,u,w $, then system \eqref{eq:system} under the controller \eqref{eq:contraction-control} is universal exponentially stable with rate $ \lambda $ and overshoot $ C=\sqrt{\frac{a_2}{a_1}} $.
\end{thm}
\begin{thm}[\cite{Manchester:2018}]
	The closed-loop system of \eqref{eq:system} and \eqref{eq:contraction-control} has an universally $ \mathcal{L}_2 $-gain bound $ \alpha $ if there exists a uniformly bounded metric $ M(x) $ such that \eqref{eq:contraction-performance-LMI} holds for all $ x,u,w $.
\end{thm}

\section{Case Study}\label{sec:case-study}

Consider the following nonlinear system
\begin{equation}\label{eq:system-example-1}
	\begin{bmatrix}
		\dot{x}_1 \\ \dot{x}_2
	\end{bmatrix}=
	\begin{bmatrix}
		-x_1-x_2+w \\ 1-e^{-x_2}+u
	\end{bmatrix}
\end{equation}
where $ w $ is a measurable reference signal. This example was used in \cite{Rugh:1991} to illustrate gain scheduling design for nonlinear systems. Assume that the control task is to stabilize the system at the equilibrium family
\begin{equation}
x_\mathrm{e}=\begin{bmatrix}
0 \\ w_\mathrm{e}
\end{bmatrix},\; u_\mathrm{e}=e^{-w_\mathrm{e}}-1.
\end{equation} 
In this section, we compare tracking control of \eqref{eq:system-example-1} with time-varying $ w_e $ using LPV and CCM-based approaches. 

For LPV design, we introduce the scheduling as $\sigma=e^{-w_\mathrm{e}}$ which in terms of the equilibrium point relation is equivalent with $\sigma=e^{-x_{2,\mathrm{e}}}$. We can obtain an LPV model of the system with coefficient matrices
\begin{equation}\label{eq:lpv-control-1}
A(\sigma)=
\begin{bmatrix}
-1 & -1 \\ 0 & \sigma
\end{bmatrix},\;
B(\sigma)=
\begin{bmatrix}
0 \\ 1
\end{bmatrix}.
\end{equation}
To derive a gain-scheduling controller, we consider placing both the closed-loop eigenvalues at $ -2 $, leading to
\begin{equation}
K(\sigma)=
\begin{bmatrix}
1 & -3-\sigma
\end{bmatrix}.
\end{equation}
Then, the control law \eqref{eq:GS-control} with the choice of $\sigma=e^{-w}$ corresponds to the nonlinear controller
\begin{equation}\label{eq:lpv-controller-1}
	u=u_e(w)+x_1-(3+e^{-w})(x_2-w)
\end{equation}
which is referred as Gain-Scheduled Controller (GSC) 1. 
The differential dynamics of the closed-loop system can be represented by
\begin{equation}
	\dot{\delta}_x=\mathcal{A}(x)\delta_x=\begin{bmatrix}
	-1 & -1 \\ 1 & a(x_2,w)
	\end{bmatrix}\delta_x
\end{equation}
where $ a(x_2,w)=e^{-x_2}-e^{-w}-3 $. Since $ \mathcal{A}(x) $ has positive eigenvalues if $ x_2<-\ln(4+e^{-w}) $, the closed-loop system is unstable in this region.

By implementing the scheduling law according to the equilibrium relation $\sigma=e^{-x_{2}}$, \eqref{eq:GS-control-implementation} gives the controller GSC 2 as follows
\begin{equation}
u=x_1+e^{-x_2}-1.
\end{equation}
Linearization of the closed-loop system at the reference $ x_\mathrm{e}(w(t))=[0,w(t)]^\top $ yields
\begin{equation}\label{eq:dd-gsc2}
\dot{x}_\delta=\begin{bmatrix}
-1 & -1 \\ 1 & 0
\end{bmatrix}x_\delta-
\begin{bmatrix}
0 \\ \dot{w}
\end{bmatrix}.
\end{equation}
The closed-loop system is globally exponential stable. But the differential dynamics have eigenvalues $ \lambda_{1,2}=-1/2\pm\sqrt{3}/2i $ with larger real parts than the specified ones $ \lambda_{1,2}=-2 $. This mismatch is caused by the hidden coupling terms:
\begin{equation}\label{eq:hct}
	K_h(\sigma)\overset{\Delta}{=}\frac{\partial u_e(\sigma)}{\partial \sigma}-K(\sigma)\frac{\partial x_e(\sigma)}{\partial \sigma}-\frac{\partial\kappa(x_e(\sigma),\sigma)}{\partial \sigma}=3.
\end{equation}

For CCM based control design, we choose the following differential state feedback control
\begin{equation}
	\delta_u=K(x)\delta_x
\end{equation}
with $ K(x)=\begin{bmatrix}
1 & -(3+e^{-x_2})
\end{bmatrix} $. This leads to exponentially stable closed-loop differential dynamics with the same eigenvalues as the LPV controller. Thus, we can obtain a constant CCM for the closed-loop system, which implies that the geodesic between $ x^* $ and $ x $ is a straight line (i.e., $ \gamma(s)=(1-s)x^*+sx $). Further, the CCM controller can be computed as
\begin{equation}\label{eq:contraction-control-1}
	\begin{split}
	u&=u^*+\int_{0}^{1}K(\gamma(s))(x-x^*)ds\\
	&=x_1+e^{-x_2}-1-3(x_2-w)+\dot{w}
	\end{split}
\end{equation}
where the target trajectory is $ x^*(t)=[0,w(t)]^\top ,\, u^*(t)=e^{-w(t)}-1+\dot{w}(t) $. The closed-loop system 
\begin{equation}
	\frac{d}{dt}
	\begin{bmatrix}
	x_1 \\ x_2-w
	\end{bmatrix}=
	\begin{bmatrix}
	-1 & -1 \\ 1 & -3
	\end{bmatrix}
	\begin{bmatrix}
	x_1 \\ x_2-w
	\end{bmatrix}
\end{equation}
is globally exponential stable at $ x^*(t) $. 

\begin{figure}[!bt]
	\centering\includegraphics[width=0.9\linewidth]{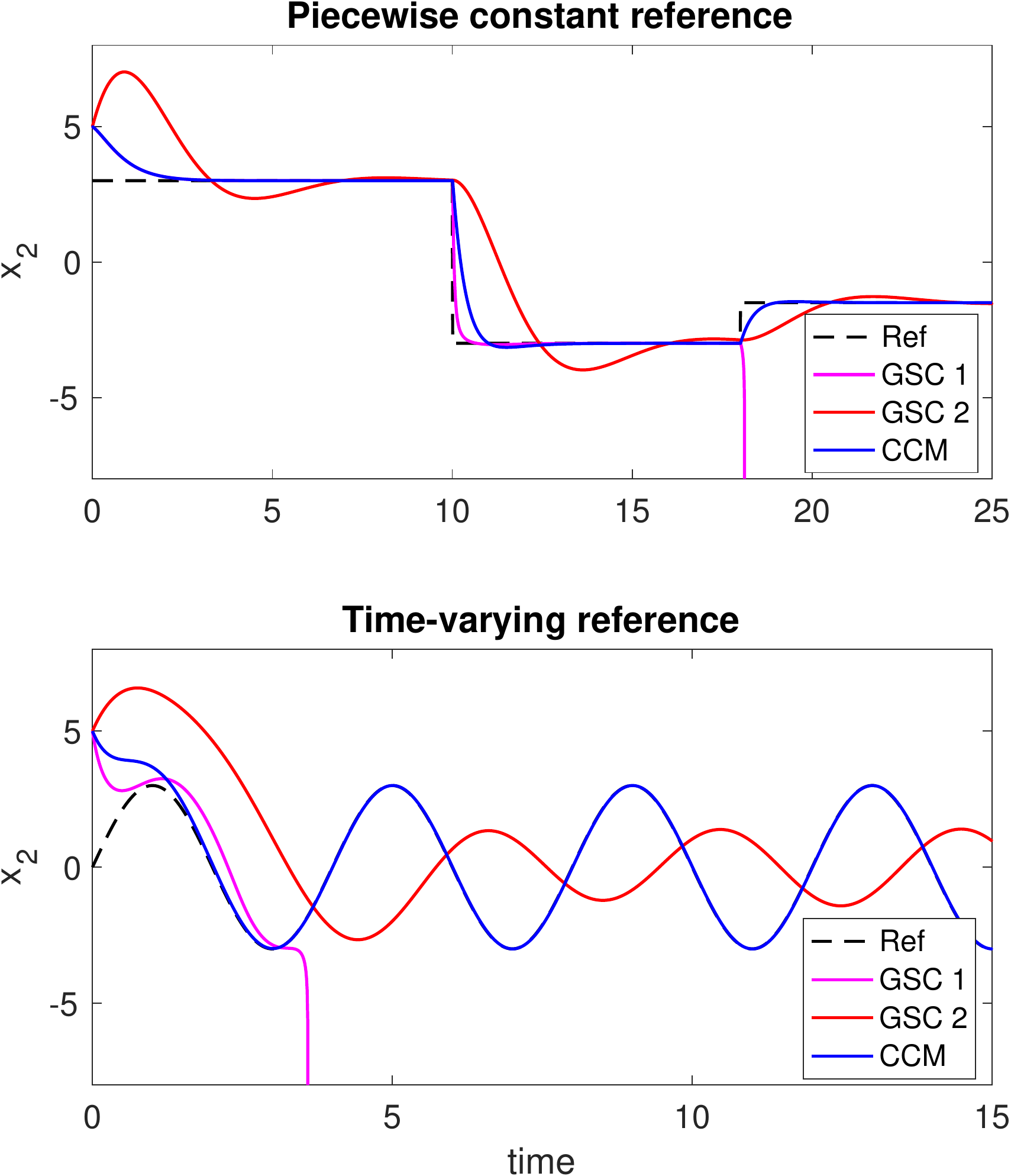}
	\caption{Closed-loop trajectories for piecewise-constant and time-varying reference.}\label{fig:cmp}
\end{figure}

For comparison study, we consider two tracking scenarios: piecewise-constant setpoints and time-varying references. As shown in Fig.~\ref{fig:cmp}, the closed-loop system under GSC 1 is not stable when the state $ x_2 $ enters into certain regions. Although the controller GSC 2 can ensure equilibrium-independent stability, the transitions are much slower and contain oscillations, compared with other controllers. This is caused by the hidden coupling terms in \eqref{eq:hct}. For time-varying references, GSC 2 cannot reach zero error due to the term $ \dot{w} $ in \eqref{eq:dd-gsc2}. The CCM based approach overcomes these issues and achieves universal stability.

\section{Conclusion}

In this paper, we investigated the apparent connection between contraction theory based nonlinear controller design and the gain-scheduling approach which corresponds to LPV control based on local linearization of the nonlinear system. We show that the CCM based control is an extended LPV gain scheduling approach as it yields a control realization without any hidden coupling term and achieves universal stability.



\bibliography{ifacconf}             


\end{document}